\documentclass[%
 reprint,aps,
 amsmath,amssymb,longbibliography,superscriptaddress]{revtex4-2}
\usepackage{booktabs}
\usepackage{tikz}
\usepackage{calc}
\usepackage{natbib}
\usepackage{booktabs}
\usepackage{longtable}
\usepackage{hyperref}
\hypersetup{
    colorlinks=true,
    linkcolor=black,
    filecolor=black,      
    urlcolor=black,
    citecolor=black,
    breaklinks
}

\usetikzlibrary{intersections}
\usetikzlibrary{decorations.pathmorphing}
\usetikzlibrary{calc}
\usetikzlibrary{shapes}

\newcommand\brak[1]{\langle #1 \rangle}
\usepackage{amsmath}

\begin{document}

\title{Singularities of massless scattering and cluster algebras}%

\author{Lara Bossinger}
\email{lara@im.unam.mx}
\affiliation{%
  Instituto de Matem\'aticas, Unidad Oaxaca, Universidad Nacional Aut\'onoma de M\'exico, Le\'on 2, 68000 Oaxaca, Mexico
}%
\author{James Drummond}
\email{j.m.drummond@soton.ac.uk}
\affiliation{%
School of Physics and Astronomy, University of Southampton, Southampton, SO15 5AT, UK
}%
\author{Ross Glew}%
\email{r.glew@herts.ac.uk}
\affiliation{%
  Department of Physics, Astronomy and Mathematics, University of Hertfordshire, Hatfield, Hertfordshire, AL10 9AB, UK
}%
\author{\"Omer G\"urdo\u gan}
\email{o.c.gurdogan@soton.ac.uk}
\affiliation{%
School of Physics and Astronomy, University of Southampton, Southampton, SO15 5AT, UK
}%
\author{Rowan Wright}%
 \email{Rowan.Wright@soton.ac.uk}
\affiliation{%
School of Physics and Astronomy, University of Southampton, Southampton, SO15 5AT, UK
}%

%\date{\today}

\begin{abstract}
Partial flag varieties arise in the context of massless scattering kinematics. They can be associated to both spinor-helicity variables and momentum twistor variables in two separate yet natural ways \cite{JMDCERN, Bossinger:2024apm, pokraka2025symbolalphabetsqcdflag}. Here we report on evidence at five and six points that the cluster algebras associated to these partial flag varieties contain information relevant to non-dual conformal massless scattering amplitudes and related observables. At five points both spinor-helicity and momentum twistor cluster structures capture similar information about symbol alphabets. At six points we demonstrate that the momentum twistor structure captures a larger subset of the alphabet compared to the spinor-helicity one. We also observe that the associated cluster structures correctly predict the appearance of certain triples of symbol letters related to cluster mutation relations. 
\end{abstract}

\maketitle
\section{Introduction}
Computation of multi-loop scattering amplitudes and related observables is one of the active frontiers of Quantum Field Theory. In a generic QFT this task typically requires the expression of the desired observable as a linear combination of Feynman integrals that one needs to evaluate. Despite significant recent advances in e.g. the technique of differential equations \cite{Henn_2013}, the computation of such integrals remains a formidable task. However, in certain special cases it is possible to produce results at very high loop orders by means of a symbol bootstrap, bypassing the computation of loop integrals. The starting point of the symbol bootstrap is a list of singularities of the observable, called the \emph{alphabet}. The success of the symbol bootstrap for e.g. six-point and seven-point amplitudes in planar Maximally Supersymmetric Yang-Mills (MSYM) \cite{Dixon_2011, Dixon_2012, Dixon_2014,Drummond_2019_fourloopheptagon}, relies on the working conjecture that the alphabet is provided by the coordinates of a Grassmannian cluster algebra \cite{fomin2001clusteralgebrasifoundations, scott2003grassmanniansclusteralgebras}. It has been a long-standing endeavour to formulate a similar principle that prescribes the alphabets of more general observables \cite{Chicherin_2021, aliaj2025exceptionalclusteralgebrahiggs}. 

It should be noted that a generic observable in a generic QFT may differ from six-point and seven-point amplitudes in planar MSYM in multiple ways. For example, generic observables lack dual conformal symmetry, require the appearance of non-rational functions in the alphabet, and require functions beyond genus-zero hyperlogarithms, such as elliptic polylogarithms.

In this note we nevertheless demonstrate that in the ``polylogarithmic world'', Grasmmannian cluster algebras, in particular their partial flag variety subalgebras, capture a large amount of the known singularities of Feynman integrals and observables whose kinematics can be embedded into those of massless scattering processes, such as scattering amplitudes, form factors, correlation functions of Wilson loop operators with local operators. We first review how the massless kinematics can be embedded into partial flag varieties in two different ways, namely through identifications with spinor-helicity variables or with momentum twistors. We then summarise the state of the art at five points where both descriptions successfully predict all physically relevant non-planar singularities but only the momentum-twistor embedding is able to make correct predictions for adjacency relations for two-loop data. 
We turn to two-loop six particle scattering, and list which singularities can be found as $\mathcal{A}$ coordinates of the partial flag varieties corresponding to the two embeddings of the kinematics in the flag variables \footnote{For six particles, both partial flag varieties turn out to be $\operatorname{F}(2,4;6)$. 
However, the two maps between cluster coordinates and kinematics are very different.}. We finally provide a survey of different observables at two loops, and the extent to which their singularities and analytic structure can be described by partial flag varieties.

\section{Momentum twistors and cluster algebras}

\subsection{Momentum twistors and cluster algebras for dual conformal scattering}

It is well-known that the kinematic space of planar scattering amplitudes  is intimately related to the Grassmannians through momentum twistors. Planarity implies a notion of ordering of the $n$ particles, which in turn allows the parametrisation of momenta $p_i$ in terms of dual coordinates
\begin{equation}
  p_i=x_{i+1}-x_{i}\,.
\end{equation}
Such variables solve the momentum conservation condition $\sum_{i=1}^{n}p_i=0$. Then one defines $n$ momentum twistors $Z_{i} \in \mathbb{CP}^3$ (as in Figure 1), which have homogeneous coordinates
\begin{equation}
  Z_i =\left(
  \begin{array}{c}
    \lambda_i^\alpha\\
    (x_i)_{\alpha\dot\alpha } \lambda_i^\alpha\,
  \end{array}
\right),
\end{equation}
where $\lambda_i$ is the holomorphic spinor in the spinor-helicity representation of a massless momentum $p_i$ as $p^\mu(\sigma_{\mu})^{\alpha \dot\alpha} = \lambda^{\alpha}\tilde \lambda^{\dot\alpha}$ and $x_{\alpha \dot\alpha} = x^\mu(\sigma_{\mu})_{\alpha \dot\alpha}$. Due to the fact that $(x_{i+1}-x_i)_{\alpha\dot\alpha}\lambda_i^\alpha = 0$, all points along the null line $x_i+t p_i$ correspond to the momentum twistor $Z_i$. Similarly the twistor line passing through $Z_{i-1}$ and $Z_{i}$, denoted as $Z_{i-1}\wedge Z_{i}$, corresponds to the dual coordinate $x_i$.

Scattering amplitudes are functions of Pl\"ucker coordinates $\brak{ijkl} = \det [Z_iZ_jZ_kZ_l]$ and two-brackets $\brak{ij} = \det [\lambda_i\lambda_j]$, which satisfy Pl\"ucker identities:
\begin{equation}
  \begin{aligned}[t]
    \brak{ijkl}Z_m + \text{cyclic}= \,& 0\\
    \brak{ij}\lambda_k + \text{cyclic}= \,& 0\,,
  \end{aligned}
\end{equation}
and they transform covariantly under
\begin{equation}
  \label{eq:littlegroup}
  \lambda_i \mapsto t_i \lambda_i ,
\end{equation}
with weights according to the helicity of the particle with momentum $p_i$.

One can strip out helicity-dependence as well as the IR divergences of the amplitude, and define a finite ``remainder function'' that is invariant under (\ref{eq:littlegroup}), and depends on ratios such as
$\frac{\brak{ijkl}}{\brak{ij}\brak{kl}}$\,
which in special cases are equal to the Mandelstam invariants
\begin{equation}
  s_{i,i+1,\dotsc,j-1}
  =
  x_{ij}^2
  =
  \frac{\brak{i-1 i j-1 j}}{\brak{i-1 i}\brak{j-1 j}}\,.
\end{equation}

Dual-conformal symmetry of amplitudes in MSYM amounts to the cancellation of two brackets, making the amplitude a function of homogeneous ratios of Pl\"ucker coordinates, i.e a function on the configuration space $\operatorname{Conf}(4,n)$. This space admits a cluster structure \cite{Scott}, with the $\mathcal{X}$-coordinates of the corresponding $\operatorname{Gr}(4,n)$ cluster algebra providing coordinates for $\operatorname{Conf}(4,n)$. In \cite{golden2014clusterpolylogarithmsscatteringamplitudes}, it was observed that the symbol letters of the known two-loop MHV amplitudes of planar MSYM are cluster coordinates. Such observations were extended to higher loops for six-point and seven-point amplitudes \cite{Dixon_2011, Dixon_2017} . Moreover, non-rational letters involving square roots can be linked to infinite mutation sequences for certain amplitudes at eight points \cite{drummond2019algebraicsingularitiesscatteringamplitudes, Henke_2020} and nine points \cite{Henke_2021}. Beyond nine points, elliptic functions appear in amplitudes at two loops \cite{Bourjaily_2018} and their relation to cluster algebras is as yet unclear. For a review of cluster algebras, we refer the reader to \cite{fomin2024introductionclusteralgebraschapters}.

\newcommand\drawtwistorline[4]{
  \draw[name path=#3] ($(#2:#1)+(#2+90:#4/2)$) -- ($(#2:#1)+(#2-90:#4/2)$);
}
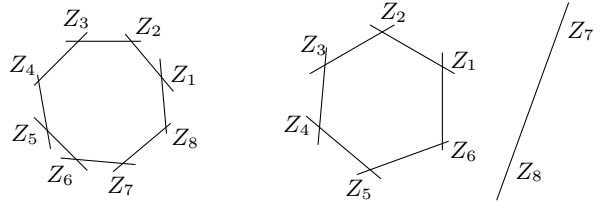
\begin{figure}
  \centering
  \begin{tikzpicture}
    \path (-1.8,-2) -- (1.8,2);
    \def\twistorsize{0.8}
    \def\twistorlinelength{1}
    \drawtwistorline{\twistorsize}{5}{line1}{\twistorlinelength}
    \drawtwistorline{\twistorsize}{40}{line2}{\twistorlinelength}
    \drawtwistorline{\twistorsize}{90}{line3}{\twistorlinelength}
    \drawtwistorline{\twistorsize}{135}{line4}{\twistorlinelength}
    \drawtwistorline{\twistorsize}{190}{line5}{\twistorlinelength}
    \drawtwistorline{\twistorsize}{225}{line6}{\twistorlinelength}
    \drawtwistorline{\twistorsize}{265}{line7}{\twistorlinelength}
    \drawtwistorline{\twistorsize}{310}{line8}{\twistorlinelength}
    
    \path [name intersections={of=line1 and line2,by=E}];
    \path ($(E) + (5:0.3)$) node {$Z_1$};
    \path [name intersections={of=line2 and line3,by=E}];
    \path ($(E) + (40:0.3)$) node {$Z_2$};
    \path [name intersections={of=line3 and line4,by=E}];
    \path ($(E) + (95:0.3)$) node {$Z_3$};
    \path [name intersections={of=line4 and line5,by=E}];
    \path ($(E) + (135:0.3)$) node {$Z_4$};
    \path [name intersections={of=line5 and line6,by=E}];
    \path ($(E) + (190:0.3)$) node {$Z_5$};
    \path [name intersections={of=line6 and line7,by=E}];
    \path ($(E) + (225:0.3)$) node {$Z_6$};
    \path [name intersections={of=line7 and line8,by=E}];
    \path ($(E) + (265:0.3)$) node {$Z_7$};
    \path [name intersections={of=line8 and line1,by=E}];
    \path ($(E) + (330:0.3)$) node {$Z_8$};
  \end{tikzpicture}
  \begin{tikzpicture}
    \path (-1.8,-2) -- (1.8,2);
    \def\twistorsize{0.8}
    \def\twistorlinelength{1.3}
    \drawtwistorline{\twistorsize}{0}{line1}{\twistorlinelength}
    \drawtwistorline{\twistorsize}{60}{line2}{\twistorlinelength}
    \drawtwistorline{\twistorsize}{120}{line3}{\twistorlinelength}
    \drawtwistorline{\twistorsize}{175}{line4}{\twistorlinelength}
    \drawtwistorline{\twistorsize}{230}{line5}{\twistorlinelength}
    \drawtwistorline{\twistorsize}{290}{line6}{\twistorlinelength}
    
    \path [name intersections={of=line1 and line2,by=E}];
    \path ($(E) + (15:0.3)$) node {$Z_1$};
    \path [name intersections={of=line2 and line3,by=E}];
    \path ($(E) + (60:0.3)$) node {$Z_2$};
    \path [name intersections={of=line3 and line4,by=E}];
    \path ($(E) + (120:0.3)$) node {$Z_3$};
    \path [name intersections={of=line4 and line5,by=E}];
    \path ($(E) + (175:0.3)$) node {$Z_4$};
    \path [name intersections={of=line5 and line6,by=E}];
    \path ($(E) + (240:0.3)$) node {$Z_5$};
    \path [name intersections={of=line6 and line1,by=E}];
    \path ($(E) + (340:0.3)$) node {$Z_6$};

    \draw[] ($(2,0)+(70:1.4)$) -- ($(2,0)-(70:1.4)$);
    \node at ($(2,0)+(70:1)+(0.3,0)$) {$Z_7$};
    \node at ($(2,0)-(70:1)+(0.3,0)$) {$Z_8$};

  \end{tikzpicture}
  
  \caption{Momentum-twistor configurations with (left) and without (right) dual conformal symmetry. Dual-conformal kinematics of $8$-point scattering is parameterised by 8 bitwistors with a particular ordering. Eight twistors can also be treated in a way where two of the twistors are interpreted as the infinity bitwistor \(I\) such that \(\langle ijI \rangle = \langle ij \rangle\).}
\end{figure}

\subsection{Adjacency rules and multiples}
In \cite{Drummond:2017ssj} it was shown that cluster algebras not only provide an alphabet for scattering amplitudes, but also, in certain cases, predict which letters are allowed to appear next to each other in their symbols. The latter statement means that double discontinuities around branch points that are not found in a cluster together must vanish, a phenomenon called \emph{cluster adjacency}. Note that all antisymmetric integrable length-2 words in cluster \(\mathcal{A}\) coordinates are necessarily cluster adjacent, and cluster adjacency is a constraint on symmetric length-2 words.

The requirement of cluster adjacency on neighbouring letters has implications on longer strings with cluster algebraic interpretations. For example, it is natural to ask what kind of triples can appear in cluster-adjacent polylogarithms. The answer (by direct computation in cases with finite cluster algebras providing rational alphabets) turns out to be interesting: namely, all triples can be identified in one of the two forms
\begin{equation}
  a\otimes b\otimes c \qquad \text{or}\qquad a \otimes X_{a a'} \otimes a'\,
  \label{tripletypes}
\end{equation}
where $\mathcal{A}$ coordinates $a,b,c$ can be found in a cluster together, $a'$ is a mutation of $a$, and $X_{a a'}$ is the cluster $\mathcal{X}$-coordinate on the node of $a$ in a cluster from which it mutates to $a'$. We emphasise that the cases of six-point and seven-point amplitudes in planar MSYM are indeed cases for which a suitable cluster adjacent finite remainder can be defined and the triples of letters found therein are indeed of the two types appearing in (\ref{tripletypes}).

In $A_n$ type cluster algebras, corresponding to Grassmannians of type $\operatorname{Gr}(2,n)$, all pairs of cluster variables are either compatible or form a mutation pair $(a,a')$. Going beyond $A_n$, for example, in an algebra of type $D_4$, corresponding to $\operatorname{Gr}(3,6)$, one finds pairs of cluster variables which are neither compatible nor mutation partners. Such letters are forbidden by cluster adjacency from appearing next to each other in the symbol and furthermore, they are forbidden from appearing separated by only one slot by (\ref{tripletypes}). In this case it turns out that they can appear in integrable words of length four, separated by two slots, and with a unique weight two integrable word separating them.

Thanks to these facts on polylogarithm symbols constructed from cluster algebra alphabets, one can probe how much a cluster algebra governs the singularity structure of a given observable beyond just the prediction of the symbol alphabet, by analysing the observed words in its symbol.

\subsection{Partial Flag Varieties}
\emph{Partial flag varieties} provide a natural generalisation of Grassmannians. Let us first recall the definition of a \emph{flag} in a finite dimensional vector space \(\mathcal{V}\) as a sequence of subspaces each contained in the next, i.e.
\[\{0\} \subset \mathcal{V}_1 \subset \mathcal{V}_2 \subset \dotsm \subset \mathcal{V}_{k+1} = \mathcal{V}.\]
Denoting the dimension of \(\mathcal{V}_i\) as \(d_i\), so that
\[0 = d_0 < d_1 < d_2 < ... < d_{k+1} = n,\]
we call the flag \emph{partial} if the sequence \(\{d_1,\dotsc,d_k\}\) is anything other than the full set of integers from \(1\) to \(n-1\) inclusive. 

We may use the dimensions themselves to label a partial flag variety, i.e. writing \(\operatorname{F}(d_1,d_2,d_3,...,d_k;n)\) for the set of all flags of length \(k\) with dimension sequence $(d_1,\dots,d_k)$. 
In particular, the Grassmannian \(\operatorname{Gr}(k,n)\) is simply the partial flag variety \(\operatorname{F}(k;n)\). 
Just as for the Grassmannian, we may associate a cluster algebra to these partial flag varieties \cite{Bossinger:2024apm} \cite{geiss2007partialflagvarietiespreprojective}, and these provide a promising starting point for generalising the application of cluster algebras to amplitudes without dual-conformal symmetry, because they provide a natural identification of the two-brackets.

Relevant to the study of \(n\)-particle massless scattering without dual conformal symmetry will be the cases of \(\operatorname{F}(2,4;n)\) and \(\operatorname{F}(2,n-2;n)\), which we review in the next section.

\subsection*{Example: \(\operatorname{F}(2,3;5) \cong D_4\)}
The cluster algebra associated to the partial flag variety \(\operatorname{F}(2,3;5)\) is of finite \(D_4\) mutation type and in fact is isomorphic to Gr\((3,6)\) \cite[Eq. 6.12]{bossinger2022adjacencyscatteringamplitudesgrobner}. 
In particular, for the Grassmannian \(\operatorname{Gr}(3,6)\) we can interpret Plücker coordinates \(p_{ijk}\) where \(i, j, k \in \{1,2,3,4,5\}\) as precisely that three-index Plücker \(p_{ijk}\) in the flag, while \(p_{ij6}\) (to which any Plücker coordinate containing \(6\) can be related by total antisymmetry) is interpreted as the two-index Plücker \(p_{ij}\) in the flag.

\begin{figure}
  \centering

 \begin{tikzpicture}
  \tikzstyle{frozen} = [draw=blue,fill=none,outer sep=2mm, inner sep=1mm];
 
    \pgfmathsetmacro\qgs{1.4}
    \node[frozen] (f0) at (-2*\qgs,\qgs) {$p_{123}$};
    \node (t1) at (-\qgs,\qgs)  {$ p_{124}$};
    \node (b1) at (-\qgs,0) {$p_{134}$};
    \node (t2) at (0,\qgs)  {$ p_{125}$};
    \node (b2) at (0,0) {$p_{145}$};
    \node[frozen] (f1) at (\qgs,\qgs)  {$ p_{12}$};
    \node[frozen] (f2) at (\qgs,0) {$p_{15}$};
    \node[frozen] (f3) at (-\qgs,-\qgs) {$p_{234}$};
    \node[frozen] (f4) at (0,-\qgs) {$p_{345}$};
    \node[frozen] (f5) at (\qgs,-\qgs) {$p_{45}$};
    \draw[-latex] (t1) -- (t2);
    \draw[-latex] (b1) -- (b2);

    \draw[-latex] (t1) -- (b1);
    \draw[-latex] (t2) -- (b2);

    \draw[-latex] (b1) -- (f3);
    \draw[-latex] (t2) -- (f1);
    \draw[-latex] (b2) -- (f2);
    \draw[-latex] (b2) -- (f4);

    % Diagonals
    \draw[-latex] (f0) -- (t1);
    \draw[-latex] (b2) -- (t1);
    \draw[-latex] (f4) -- (b1);
    \draw[-latex] (f5) -- (b2);
    
\end{tikzpicture}
\caption{The initial cluster of $\operatorname{Gr}(3,6)$, which is isomorphic to the flag $\operatorname{F}(2,3;5)$. Coordinates are labelled according to the $\operatorname{F}(2,3;5)$ convention.}

\end{figure}
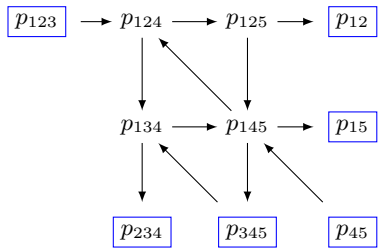
The initial cluster, depicted in Figure 2, contains active coordinates
\begin{equation}
\{p_{124}, p_{125}, p_{134}, p_{145} \}
\end{equation}
while the frozen coordinates are
\begin{equation}
\{p_{123}, p_{234}, p_{345}, p_{12}, p_{15}, p_{45}\}.
\end{equation}
Performing all possible sequences of mutations, one finds a total of sixteen active coordinates and thus twenty-two \(\mathcal{A}\)-coordinates in total, distributed across fifty distinct clusters. These are the twenty Plücker coordinates of the form \(p_{ijk}\) with \(i < j < k\) drawn from the labels \(1\) to \(6\) (and interpreted appropriately in the flag dependent on whether \(6\) is present), as well as the two quadratics 
\begin{align}
p_{124}\, p_{356} - p_{123}\, p_{456} &\cong p_{124}\, p_{35} - p_{123}\, p_{45} \notag \\
p_{235}\, p_{145} - p_{234}\, p_{156} &\cong p_{235}\, p_{145} - p_{234}\, p_{15}.
\end{align}

We will return to consider this cluster algebra when reviewing the spinor helicity cluster structure for five particle massless scattering. 

\subsection*{Example: \(\operatorname{F}(2,4;5) \cong A_4\)}
\begin{figure}
  \centering
  \begin{tikzpicture}
      \tikzstyle{frozen} = [draw=blue,fill=none,outer sep=2mm, inner sep=1mm];
    \pgfmathsetmacro{\qgs}{1.4}
    \node (t1) at (0,\qgs) {$p_{1235}$};
    \node (m1) at (0,0) {$q_1$};
    \node[frozen] (ffour) at (-\qgs,0) {$p_{12}$};
    \node (b2) at (0,-\qgs) {$p_{1345}$};
    \node[frozen] (f0) at (-\qgs,\qgs) {$p_{1234}$};
    \node[frozen] (f1) at (\qgs,\qgs) {$p_{15}$};
    \node[frozen] (f2) at (\qgs,0) {$p_{45}$};
    \node[frozen] (f3) at (2*\qgs,-\qgs) {$p_{2345}$};
    \node[frozen] (b3) at (\qgs,-\qgs) {$q_2$};
    \node (b1) at (-\qgs,-\qgs) {$p_{1245}$};

    \draw[-latex] (t1) -- (m1);
    \draw[-latex] (f0) -- (t1);
    \draw[-latex] (f1) -- (t1);
    \draw[-latex] (f2) -- (m1);
    \draw[-latex] (b2) to[out=-20,in=-160] (f3);
    \draw[-latex] (b2) -- (m1);
    \draw[-latex] (b1) -- (b2);
    \draw[-latex] (b1) -- (ffour);
    \draw[-latex] (ffour) -- (m1);
    \draw[-latex] (m1) -- (f1);
    \draw[-latex] (m1) -- (b1);

    \draw[-latex] (b3) -- (m1);
    \draw[-latex] (b3) -- (b2);
    \draw[-latex] (b3) -- (f2);

      \node (qq) at (4.5,0) {\begin{minipage}{4cm}
      \begin{equation*}     \begin{aligned}[t]
                              q_1=& p_{15}\,p_{1234} - p_{14}\,p_{1235}      \\
                              q_2=&p_{35}\,p_{1234}\,  - p_{34}\,p_{1235}      \\
                                  \end{aligned}\end{equation*}
    \end{minipage}};

  \end{tikzpicture}

\caption{A choice of initial cluster for the flag \(\operatorname{F}(2,4;5)\), which can be embedded in \(\operatorname{Gr}(4,7)\) by a sequence of mutations and freezings.}
  
\end{figure}
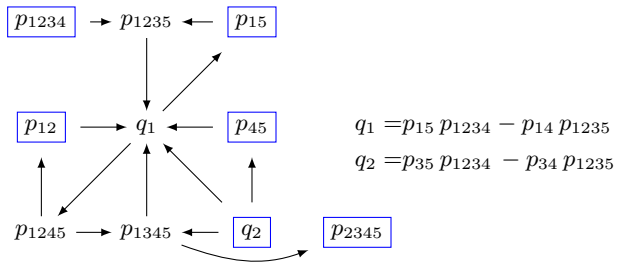
The partial flag variety \(\operatorname{F}(2,4;5)\), whose initial cluster is depicted in Figure 3, contains four active nodes and six frozen nodes. The initial cluster has coordinates 
\begin{equation}
\{a_1,a_2,a_4,a_4\} = \{p_{15}\, p_{1234} - p_{14}\, p_{1235}, p_{1345}, p_{1235}, p_{1245}\}
\end{equation}
at the active nodes, while the frozen coordinates are given by 
\begin{equation}
\{p_{2345},p_{34}\, p_{1235}-p_{35}\, p_{1234},p_{45},p_{15},p_{1234},p_{12}\}.
\end{equation}
Note that this may be realised as a subalgebra with two fewer mutable cluster variables of the Grassmannian cluster algebra \(\operatorname{Gr}(4,7)\) by performing an appropriate sequence of mutations and two freezings from the initial cluster, leading to the cluster shown in Figure 3. This ensures the twistor labels \(6\) and \(7\) appear either together in a given Plücker coordinate, or not at all. We then interpet \(p_{ijkl}\) with \(i,j,k,l \in \{1,2,3,4,5\}\) as that same Plücker coordinate \(p_{ijkl}\) in the flag, while \(p_{ij67}\) is identified with the two-index Plücker coordinate \(p_{ij}\).

This cluster algebra is of finite, \(A_4\) mutation type, and by performing all possible mutations one finds a total of \(20\) \(\mathcal{A}\)-coordinates contained within \(42\) different clusters: the active coordinates are 
\begin{multline}
\{p_{13}, p_{14}, p_{23}, p_{24}, p_{25}, p_{34}, p_{35}, p_{1235}, p_{1345}, p_{1245},\\ p_{14}\, p_{1235} - p_{15}\, p_{1234}, p_{35}\, p_{1245}-p_{45}\, p_{1235},\\ p_{34}\, p_{1245}-p_{45}\, p_{1234},p_{24}\, p_{1235}-p_{25}\, p_{1234}\}.
\end{multline}

This cluster algebra is of relevance when considering the momentum twistor cluster structure for five particle massless scattering, which we will discuss in the next section. 

\subsection*{Example: \(\operatorname{F}(2,4;6) \cong D_6^{(1)}\)}
\(\operatorname{F}(2,4;6)\) is of \(D_6^{(1)}\) mutation type, and is therefore infinite, though as a codimension two subalgebra of \(\operatorname{Gr}(4,8)\) it is of finite mutation type. The active coordinates of the initial cluster are given by 
\begin{equation}
\{p_{15}\, p_{1236}-p_{16}\, p_{1235}, p_{1456}, p_{1345},p_{1256},p_{1245},p_{1236},p_{1235}\}
\end{equation}
with frozen coordinates
\begin{multline}
\{p_{2345},p_{3456},p_{56}\, p_{1234}-p_{46}\, p_{1235} + p_{45}\, p_{1236},\\ p_{56}, p_{16}, p_{12}, p_{1234}\}.
\end{multline}

\begin{figure}
  \centering
\begin{tikzpicture}
  \tikzstyle{frozen} = [draw=blue,fill=none,outer sep=2mm, inner sep=1mm];
    \draw[line width=2mm, orange!20!white, -latex] (1,0.5)  -- (4,0.5);
    \draw[line width=2mm, orange!20!white, -latex] (4,-0.5)  -- (1.5,-2.5);

  \pgfmathsetmacro\qgs{1.5}
  \node (t1) at (-\qgs,\qgs)  {$ p_{1235}$};
  \node (m1) at (-\qgs,0)  {$p_{1245}$};
  \node (b1) at (-\qgs,-\qgs) {$p_{1345}$};
  \node (t2) at (0,\qgs) {$p_{1236}$};
  \node (m2) at (0,0) {$p_{1256}$};
  \node (b2) at (0,-\qgs) {$p_{1456}$};
  \node (t3) at (\qgs,\qgs) {$p_{1237}$};
  \node (m3) at (\qgs,0) {$p_{1267}$};
  \node (b3) at (\qgs,-\qgs) {$p_{1567}$};

  \draw[-latex] (t1) -- (t2);
  \draw[-latex] (t2) -- (t3);
  \draw[-latex] (m1) -- (m2);
  \draw[-latex] (m2) -- (m3);
  \draw[-latex] (b1) -- (b2);
  \draw[-latex] (b2) -- (b3);

  \draw[-latex] (t1) -- (m1);
  \draw[-latex] (t2) -- (m2);
  \draw[-latex] (t3) -- (m3);
  \draw[-latex] (m1) -- (b1);
  \draw[-latex] (m2) -- (b2);
  \draw[-latex] (m3) -- (b3);

  \draw[-latex] (m2) -- (t1);
  \draw[-latex] (m3) -- (t2);
  \draw[-latex] (b2) -- (m1);
  \draw[-latex] (b3) -- (m2);

  \tikzset{shift={(4.5,0)}}
  \node (t1) at (-\qgs,\qgs)  {$ p_{1235}$};
  \node (m1) at (-\qgs,0)  {$p_{1245}$};
  \node (b1) at (-\qgs,-\qgs) {$p_{1345}$};
  \node (t2) at (0,\qgs) {$p_{1236}$};
  \node (m2) at (0,0) {$p_{1256}$};
  \node (b2) at (0,-\qgs) {$p_{1456}$};
  \node (t3) at (\qgs,\qgs) {$p_{1237}$};
  \node (m3) at (\qgs,0) {$q_1$};
  \node (b3) at (\qgs,-\qgs) {$p_{1567}$};

  \draw[-latex] (t1) -- (t2);
%  \draw[-latex] (t2) -- (t3);
  \draw[-latex] (m1) -- (m2);
  \draw[-latex] (m3) -- (m2);
  \draw[-latex] (b1) -- (b2);
  \draw[-latex] (b2) -- (b3);

  \draw[-latex] (t1) -- (m1);
  \draw[-latex] (m3) -- (t3);
  \draw[-latex] (b3) -- (m3);

  \draw[-latex] (m1) -- (b1);
  \draw[-latex] (m2) -- (b2);

  \draw[-latex] (t3) to[out=-70,in=70] (b3);

  \draw[-latex] (m2) -- (t1);
  \draw[-latex] (t2) -- (m3);
  \draw[-latex] (b2) -- (m1);
 % \draw[-latex] (b3) -- (m2);

  \tikzset{shift={(-4.5,-4.5)}}
  \node (t1) at (-\qgs,\qgs)  {$ p_{1235}$};
  \node (m1) at (-\qgs,0)  {$p_{1245}$};
  \node (b1) at (-\qgs,-\qgs) {$p_{1345}$};
  \node (t2) at (0,\qgs) {$p_{1236}$};
  \node (m2) at (0,0) {$p_{1256}$};
  \node (b2) at (0,-\qgs) {$p_{1456}$};
  \node[frozen] (t3) at (\qgs,\qgs) {$p_{1237}$};
  \node (m3) at (\qgs,0) {$q_1$};
  \node[frozen] (b3) at (\qgs,-\qgs) {$q_2$};

  \draw[-latex] (t1) -- (t2);
%  \draw[-latex] (t2) -- (t3);
  \draw[-latex] (m1) -- (m2);
  \draw[-latex] (m3) -- (m2);
  \draw[-latex] (b1) -- (b2);
  \draw[-latex] (b2) -- (m3);
  \draw[-latex] (b3) -- (b2);

  \draw[-latex] (t1) -- (m1);
%  \draw[-latex] (m3) -- (t3);
  \draw[-latex] (m3) -- (b3);

  \draw[-latex] (m1) -- (b1);
  \draw[-latex] (m2) -- (b2);

  \draw[-latex] (b3) to[out=70,in=-70] (t3);

  \draw[-latex] (m2) -- (t1);
  \draw[-latex] (t2) -- (m3);
  \draw[-latex] (b2) -- (m1);
 % \draw[-latex] (b3) -- (m2);

  \node (qq) at (4.5,0) {\begin{minipage}{4cm}
      \begin{equation*}     \begin{aligned}[t]
                                  q_1=&p_{1235}\, p_{1678} - p_{1236}\, p_{1578}     \\
                              q_2=&p_{1236}\, p_{4578} \begin{aligned}[t]
                                                       &- p_{1235}\, p_{4678}\\
                                                       &+ p_{1234}\, p_{5678}
                                                       \end{aligned}\\
                                  \end{aligned}\end{equation*}
    \end{minipage}};
\end{tikzpicture}
\caption{The flag \(\operatorname{F}(2,4;6)\) realised as a codimension two subalgebra of \(\operatorname{Gr}(4,8)\).  By freezing $q_1$, $p_{1237}$ becomes ``sequestered'', and the rest of the quiver is precisely a quiver that can be found in $\operatorname{F}(2,4;6)$}
\end{figure}
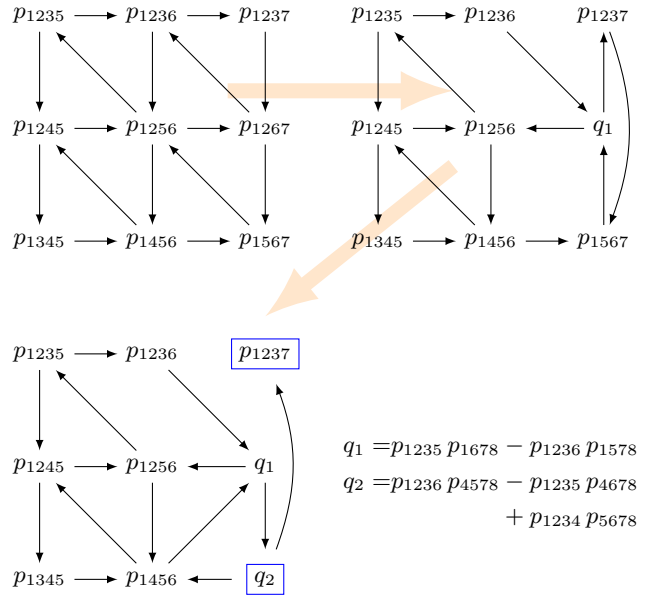

By direct analogy with the embedding of \(\operatorname{F}(2,4;5)\) inside \(\operatorname{Gr}(4,7)\), \(\operatorname{F}(2,4;6)\) may be realised as a codimension two subalgebra of the Grassmannian \(\operatorname{Gr}(4,8)\) by performing a sequence of freezings and mutations so as to ensure the twistor indices \(7\) and \(8\) always appear either together or not at all. This procedure is shown in Figure 4. A Grassmannian Plücker coordinate \(p_{ijkl}\) with \(i,j,k,l \in \{1,2,3,4,5,6\}\) is then interpreted as that same Plücker in the flag, while \(p_{ij78}\) is identified with \(p_{ij}\).

We will later see that \(\operatorname{F}(2,4;6)\) is relevant for both the momentum twistor and spinor helicity cluster structures associated to six-particle massless scattering, and moreover that its identification as a subalgebra of \(\operatorname{Gr}(4,8)\) provides guidance when truncating the infinite list of \(\mathcal{A}\)-coordinates generated by this cluster algebra. 

\section{Cluster structures associated to general massless scattering}

\subsection{Momentum twistor cluster structure in \(\operatorname{F}(2,4;n)\)}
It is straightforward to map non-dual conformal scattering kinematics for \(n\)-particle massless scattering into the flag \(\operatorname{F}(2,4;n)\), and vice versa, where we interpret \(p_{ijkl}\) as a momentum twistor \(\langle ijkl \rangle \) in the normal sense and \(p_{ij}\) as an angle bracket \(\langle ij \rangle\). This gives what we refer to as the momentum twistor cluster structure. 

Note that we may express square brackets in terms of momentum twistors and angle brackets and thus in terms of Plücker coordinates \(p_{ijkl}\) and \(p_{ij}\) by using the relation for Mandelstam invariants
\begin{equation}
\langle ij \rangle [ji] = s_{ij}
\end{equation}
along with the identity
\[s_{k,k+1, ... , l-1} = x_{k,l}^2 = \frac{\langle k-1,k,l-1,l\rangle}{\langle k-1 k \rangle \langle l-1 l \rangle}\]
to rewrite them in terms of the flag variables. For example, if we seek to express \([13]\) for six-particle scattering in terms of Plücker coordinates for the flag we have
\begin{equation}
\langle 31 \rangle [13] = s_{13} 
\end{equation}
which may be rewritten as 
\begin{equation}
\langle 31 \rangle [13] = s_{123} - s_{12} - s_{23}.
\end{equation}
So,
\begin{equation}
[13] = \frac{1}{p_{31}} \bigl(\frac{p_{6134}}{p_{61}\, p_{34}} - \frac{p_{6123}}{p_{61}\, p_{23}} - \frac{p_{1234}}{p_{12}\, p_{34}}\bigr).
\end{equation}

The map in the other direction, from four-index Plücker coordinates to square and angle brackets, is similarly straightforward. Since we concern ourselves here with the scattering of at most six particles, we can simplify the considerations by noting that at least two of the indices \(i,j,k,l\) in \(p_{ijkl}\) must be consecutive, and thus we may restrict our considerations to Plücker coordinates of the form \(p_{i-1,i,j,k}\).

Using the general result
\begin{multline}
    \langle i j k l \rangle = \langle i \ j \rangle \left[\mu_l \ \mu_k \right] + \langle i \ k \rangle \left[\mu_j \ \mu_l \right] + \langle i \ l \rangle \left[\mu_k \ \mu_j \right] \\
    + \langle k \ l \rangle \left[\mu_j \ \mu_i \right] + \langle k \ j \rangle \left[\mu_i \ \mu_l \right] + \langle l \ j \rangle \left[\mu_k \ \mu_i \right]
\end{multline}

alongside the incidence relation 
\begin{equation}
    |\mu_i] = x_i|i\rangle = x_{i+1}|i\rangle,
\end{equation}
and the Schouten identity, we have straightforwardly that
\begin{equation}
       p_{i-1,i,j,k} = \langle i-1 \ i \rangle \langle k|x_{ki}x_{ji}|j\rangle.
\end{equation}
Here we have defined \(x_{ab} \equiv x_a - x_b\). Then, using that
\begin{equation}
    x_{ab} = \sum_{r=a}^{b-1} p_r
%\end{equation}
%\begin{equation}
    = \sum_{r=a}^{b-1} |r\rangle[r|
\end{equation}
where the sums are taken to run modulo the number of particles, our expression becomes
\begin{equation}
    p_{i-1,i,j,k} = \langle i-1 \  i \rangle \sum_{q=j}^{i-1} \sum_{r=k}^{i-1} \langle k \ r \rangle[r \ q]\langle q \ j\rangle.
\end{equation}

We may thus map between the flag variables and expressions for symbol letters written in terms of Mandelstam invariants or, more generally, of spinor brackets.

\subsection{Spinor helicity cluster structure in \(\operatorname{F}(2,n-2;n)\)}

Alternatively, we may embed the kinematics in \(\operatorname{F}(2,n-2;n)\), where \(p_{ij}\) is again interpreted as an angle bracket while Plücker coordinates with \(n-2\) indices, i.e. all labels except for \(i\) and \(j\) with \(i<j\), which we denote as \({p_{\overline{ij}}}\) are identified as 

\begin{equation}
{p_{\overline{{ij}}}} = (-1)^{j-i-1}[ij].
\end{equation}
For instance, in the case \(n=5\) we have
\begin{equation}
p_{124} = p_{\overline{35}} = (-1)^{5-3-1}[35] = -[35].
\end{equation}

As discussed in the case of five points in \cite{bossinger2022adjacencyscatteringamplitudesgrobner} it is then immediate that the Schouten identities and momentum conservation, the identities constraining the spinor helicity variables, are implemented by the Plücker relations. For instance, for \(\operatorname{Gr}(3,6)\), we have three- and four-term Plücker relations. Those involving the label \(6\) in one of the two Plücker coordinates for each product, e.g.
\begin{align}
p_{123}\, p_{146}+p_{126}\, p_{134}-p_{124}\, p_{136} &= 0, \\ \notag 
p_{123}\, p_{456}-p_{156}\, p_{234}+p_{146}\, p_{235} - p_{145}\, p_{236} &= 0 
\end{align}
reduce to the statement of momentum conservation when written in terms of the kinematics.

% For example, the above become
% \begin{align}
% [45]\langle 14\rangle+ [25] \langle 12 \rangle + [35] \langle 13 \rangle &= 0, \\ \notag 
% [45] \langle 45 \rangle +  [15] \langle 15 \rangle + [14] \langle 14 \rangle - [23] \langle 23 \rangle &= 0.
% \end{align}
Similarly, Plücker relations without a six e.g. 
\begin{equation}
p_{123}\, p_{145}+p_{125}\, p_{134}-p_{124}\, p_{135} = 0
\end{equation}
and Plücker relations with a six in both Plücker coordinates for each product e.g.
\begin{equation}
p_{236}\, p_{456}+p_{256}\, p_{346}-p_{246}\, p_{356} = 0
\end{equation}
become the Schouten identities for square and angle brackets respectively. 

\section{Partial flag varieties for five-point massless scattering}

\subsection{Two loop, five-point alphabet from partial flag varieties}

The planar pentagon alphabet relevant for five-particle massless scattering at two loops consists of \(26\) symbol letters; using the notation of \cite{Chicherin_2018}, the symbol letters are (here we write \(V\) instead of \(W\) to avoid confusion with the planar \emph{hexagon} alphabet which we will discuss later)
\begin{equation}
\{V_i\}_{i=1}^{20} \cup \{V_{i}\}_{i=26}^{31}.
\end{equation}

The full non-planar alphabet at two loops is given by the planar alphabet supplemented with the letters \(\{V_i\}_{i=21}^{25}\).

Notably, \(V_{31}\) has been absent from all appropriately defined finite two-loop quantities calculated to date \cite{Abreu_2019sg} \cite{Abreu_2019sym} \cite{chicherin2022pentagonwilsonlooplagrangian} \cite{Abreu_2019}. Here we review the recovery of the symbol letters using both \(\operatorname{F}(2,n-4;n)\) and \(\operatorname{F}(2,4;n)\) in the case \(n=5\).

\subsubsection{Alphabet from $\operatorname{F}(2,3;5)$}
As we previously noted, the partial flag variety \(\operatorname{F}(2,n-2;n)\) in the case \(n=5\) is \(\operatorname{F}(2,3;5)\) which is isomorphic to the \(\operatorname{Gr}(3;6)\) cluster algebra. This finite cluster algebra of \(D_4\) mutation type endows us with a total of \(22\) cluster variables: \(20\) minors of the form \(p_{ijk}\) with \(i,j,k \in \{1,2,3,4,5,6\}\) (and where we interpet \(p_{ij6}\) as \(p_{ij}\) inside the flag), plus the two quadratics identified earlier.

Taking multiplicative combinations of the cluster variables of \(\operatorname{F}(2,3;5)\) and interpreting these as spinor helicity variables in the way described, we find that we are able to recover the letters $
V_1$ -- $V_5$, $V_{13}$, $V_{14}$, $V_{16}$ -- $V_{20}$, $V_{26}$.

In particular, we have
\begin{subequations}
  \begin{align}
    V_1 &= p_{12}\,p_{345}\qquad \text{(and its 4 cyclic copies)} \\
    V_{13} &= p_{35}\,p_{124} - p_{45}\,p_{123} \\
    V_{14} &= p_{14}\,p_{235} - p_{15}\,p_{234}  \\
    V_{16} &= p_{13}\,p_{245} \qquad \text{(and its 4 cyclic copies)}\\
    V_{26} &= \frac{p_{12}\,p_{45}\,p_{135}\,p_{234}}{p_{15}\,p_{24}\,p_{123}\,p_{345}} \qquad \text{(and its 4 cyclic copies)}
  \end{align}
\end{subequations}
Completing these symbol letters under cyclic permutations on the particle labels enables recovery of all planar letters except for \(V_{31}\) and the cyclic class \(\{V_i\}_{i=6}^{10}\). Upon completing under \emph{all} permutations we recover not only \(\{V_i\}_{i=6}^{10}\) but also the non-planar letters \(\{V_i\}_{i=21}^{25}\). 

\subsubsection{Alphabet from $\operatorname{F}(2,4;5)$}
We may alternatively investigate the embedding of five particle scattering in the partial flag \(\operatorname{F}(2,4;n)\), which in the present case is \(\operatorname{F}(2,4;5)\), a cluster algebra of finite \(A_4\) mutation type with \(20\) cluster coordinates distributed across \(42\) clusters.

Interpreting two-index Pl\"ucker coordinates as angle brackets and four-index Pl\"ucker coordinates as momentum twistors, we find by taking multiplicative combinations of the cluster coordinates that we are able to produce the following 11 symbol letters: $V_1$ -- $V_5$, $V_{12}$, $V_{15}$, $V_{17}$, $V_{19}$, $V_{20}$, $V_{26}$.

In particular, we have
\begin{subequations}
  \begin{align}
    V_1 &= \frac{p_{1235}}{p_{15}\, p_{23}}\qquad \text{(and 4 cyclic copies)}\\
V_{12} &= \frac{ p_{34}\, p_{1245} -p_{45}\, p_{1234}}{
 p_{12}\, p_{34} p_{45}}\\
V_{15} &= \frac{ 
p_{24}\, p_{1235}-p_{25}\, p_{1234}}{p_{12}\, p_{23} p_{45}}\\
V_{17} &= \frac{p_{24} ( p_{34}\, p_{1235}-p_{35}\, p_{1234} )}{
 p_{12}\, p_{23} p_{34}\, p_{45}}\\
V_{19} &= \frac{
 p_{14} (p_{35}\, p_{1245}-p_{45}\, p_{1235})}{
 p_{12}\, p_{15} p_{34}\, p_{45}}\\
 V_{20} &= 
  \frac{p_{25}(p_{14}\, p_{1235}-p_{15}\, p_{1234})}{
 p_{12}\, p_{15} p_{23}\, p_{4
   5}}\\
V_{26} &= \frac{p_{124
    5}(p_{34}\, p_{1235}-p_{35}\, p_{1234}) }{p_{24}\, p_{1235} p_{1345})}.
\end{align}
\end{subequations}
This clearly covers precisely the same cyclic classes which were obtained under the spinor-helicity embedding of the kinematics, and so under completion under cyclic permutations or all permutations, we recover the same symbol letters as in that case. 

Note that the need to use permutations to generate many symbol letters using either embedding of the kinematics is unsurprising in light of the fact that the relations between Grassmannian and flag coordinates e.g.
\begin{equation}
p_{ij6} \cong p_{ij}
\end{equation}
\begin{equation}
p_{ij67} \cong p_{ij}
\end{equation}
break the natural cyclic symmetry of the Grassmannians in which the partial flag varieties may be embedded. Although it is convenient to think of permutations as acting on the symbol letters (which are themselves naturally expressed in terms of particle momenta, not twistor labels), it is of course more natural to consider that the corresponding \emph{inverse} permutations act on the cluster variables we obtain from the canonical orientation of the flag, and therefore that permutations are required to account for contributions from cluster variables originating in \emph{different} orientations of the flag.

\subsection{Comments on five point adjacency relations}

\subsubsection{Adjacency predictions from \(\operatorname{F}(2,4;5)\)}

Given that the \(\operatorname{F}(2,3;5)\) and \(\operatorname{F}(2,4;5)\) cluster algebras are different, but enjoy identical success in generating the symbol alphabet for five point massless scattering, it is interesting to investigate to what extent either cluster algebra can make predictions about adjacency relations at five points.

Written in terms of spinor brackets, letters $V_6$ to $V_{21}$ from the non-planar two-loop five-point alphabet are quadratic. The remaining letters of the alphabet factorise into spinor brackets. It is observed empirically that two distinct quadratic letters never appear next to each other in any two-loop data computed thus far \cite{Abreu_2019sg,Abreu_2019sym,chicherin2022pentagonwilsonlooplagrangian,Abreu_2019}\, though these letters may sometimes appear separated by one. This suggests they might be cluster incompatible letters related to each other by mutation. For instance, in the coefficient of \(r_1\) in the planar pentagonal MSYM Wilson loop with Lagrangian operator insertion \cite{chicherin2022pentagonwilsonlooplagrangian}, all the terms which have different quadratic letters separated by one slot in the symbol can be combined into the following symbol terms:
\begin{multline}
2(\frac{V_1}{V_4}\otimes V_{11} \otimes \frac{V_1V_{18}}{V_3V_4} \otimes V_{13} ) + \frac{V_2}{V_4} \otimes V_{14} \otimes \frac{V_4V_{19}}{V_2V_5} \otimes V_7\\
+ \frac{V_4}{V_1} \otimes V_{11} \otimes \frac{V_1V_{16}}{V_2V_4} \otimes V_9 + \frac{V_4}{V_2}\otimes V_{14} \otimes \frac{V_2V_{16}}{V_1V_4} \otimes V_9\\
+ \frac{V_2}{V_5} \otimes V_{12} \otimes \frac{V_5V_{19}}{V_2V_4} \otimes V_7 + \frac{V_5}{V_2} \otimes V_{12} \otimes \frac{V_5V_{17}}{V_2V_3} \otimes V_{15}
\end{multline}
  
Notably there is a unique "\(X\)" between each quadratic pair, e.g. in the last term the "\(X\)" between $V_{12}$ and $V_{15}$ is $\frac{V_5V_{17}}{V_2V_3}$. By analogy with the case of planar MSYM amplitudes, we would expect that in the cluster algebra one of the cluster variables which are factors of \(V_{12}\) mutates to one of the cluster variables which are factors of \(V_{15}\), and that the \(X\) which sits between them is the \(\mathcal{X}\)-coordinate of \(V_{12}\)'s factor computed in the cluster where it mutates to \(V_{15}\)'s. 

This cannot be realised in \(\operatorname{F}(2,3;5)\) using the spinor helicity cluster structure, where in every permutation the quadratic letters not only contain cluster incompatible factors but in fact these factors do not even mutate to each other. However, in the momentum twistor cluster structure the letters $V_{12}$ and $V_{15}$ are related to the following \(\operatorname{F}(2,4;5)\) letters:
  
\begin{equation}
p_{34}\, p_{1245} -p_{45}\, p_{1234} = p_{12}\, p_{34} p_{45}V_{12}
\end{equation}
\begin{equation}
p_{24}\, p_{1235} - p_{25}\, p_{1234} = p_{12}\, p_{23} p_{45}V_{15}
\end{equation}
  
These two letters form a mutation pair in the cluster algebra, and the \(\mathcal{X}\)-coordinate of the former in the cluster where it mutates to the latter matches exactly to \(\frac{V_5V_{19}}{V_2V_4}\) when interpreted in the momentum twistor embedding.
  
The other terms listed above where quadratics are separated by one are then obtainable by performing appropriate permutations on particle labels.

\subsubsection{Classifying pairs of quadratic letters}

Let us consider in more detail the classes of quadratic letters which may appear separated in the same word. 

Each quadratic letter is of the form
\begin{equation}
Q_{i,j;k,l;m} = s_{i,j} - s_{k,l}
\end{equation}
for \(\{i,j,k,l,m\} = \{1,2,3,4,5\}\).
Note that we indeed have \(15\) quadratics under permutation completion since these enjoy the symmetry (up to an irrelevant sign)
\begin{equation}
Q_{i,j;k,l;m} = Q_{j,i;k,l,m} = Q_{i,j;l,k;m} = Q_{j,i;l,k;m}.
\end{equation}

There are clearly three different distinct permutation classes of pairs of quadratics: 

\begin{enumerate}
\item \((Q_{i,j;k,l;m}, Q_{i,j;k,m;l})\). These are the pairs of quadratics which appear separated by one slot in presently computed two-loop planar data, e.g. the pentagonal Wilson loop with Lagrangian operator insertion \cite{chicherin2022pentagonwilsonlooplagrangian}, and planar QCD \cite{Abreu_2019}. Such triples are also seen in non-planar data e.g. \(\mathcal{N}=8\) supergravity \cite{Abreu_2019sg}, non-planar MSYM \cite{Abreu_2019sym}, including variations using non-planar letters such as
\begin{equation}
V_{6} \otimes \frac{V_1V_{17}}{V_{18}V_{19}} \otimes V_{24}
\end{equation}
and
\begin{equation}
V_{21} \otimes \frac{V_2V_{19}}{V_{18}V_{20}} \otimes V_{25}.
\end{equation}
These are all permutations of the triple given in the last section which has an interpretation in \(\operatorname{F}(2,4;5)\). 

\item \((Q_{i,j;k,l;m}, Q_{i,k;j,l;m})\). These appear separated by one slot in non-planar data, with a unique combination sitting between. For instance, 
\begin{equation}
V_{6} \otimes \frac{V_3V_{19}}{V_4V_{17}} \otimes V_{12}
\end{equation}
is a triple observed in both non-planar MSYM \cite{Abreu_2019sym} and \(\mathcal{N}=8\) supergravity \cite{Abreu_2019sg}. This does \emph{not} have an interpretation inside \(\operatorname{F}(2,4;5)\), where there is not a permutation under which \(V_{6}\) and \(V_{12}\) are simultaneously identifiable. However, the uniqueness of the combination in between the quadratic letters is highly indicative of some underlying cluster structure. Indeed, it is interesting to observe that the letters involved even enjoy the exchange-type relation
\begin{equation}
V_6V_{12} = V_3V_{19}-V_{4}V_{17} 
\end{equation}
which, based on the cluster mutation relation, would be expected if \(\frac{V_3V_{19}}{V_4V_{17}}\) is to be realised as an \(\mathcal{X}\)-coordinate in a cluster of some cluster algebra where \(V_6\) mutates to \(V_{12}\). Note that the relative sign on the right-hand sign, which would crucially be a \(+\) in the cluster exchange relation to ensure positivity, is arbitrary here since symbol letters are dlogs and thus only defined up to a sign.

\item \((Q_{i,j;k,l;m}, Q_{i,k;j,m;l})\) does not appear in any two-loop data, but strikingly this is precisely the type of pair of quadratics mutually identifiable under some permutation inside \(\operatorname{F}(2,3;5)\) using the spinor helicity embedding of the kinematics. Since we would expect such quadratics to be separated by at least \emph{two} slots in the symbol, and we only have access to weight four data (with quadratics forbidden from appearing in the first entry), it may be that these contributions from \(\operatorname{F}(2,3;5)\) will begin to appear in higher loop data. 
\end{enumerate}

\subsubsection{Adjacency relations in integrable triples and quadruples and beyond}
Since the available two-loop data is only of weight four, with stringent restrictions on the first entry and (conjecturally) the second entry, it is instructive to probe the structure of integrable triples (or quadruples) more generally. For instance, it is a straightforward calculation to compute all weight three (or four) symbols constructed from the non-planar two-loop pentagon alphabet which obey the constraint that no two quadratic letters may appear adjacently.

Doing so, quadratic pairs of Type 1 and Type 2 arise with only one unique \(X\) seen between them. Note that, a priori, there was no need for the inhabitant of the slot separating these quadratics to be unique in this calculation, and certainly if one imposes arbitrary adjacency restrictions on a symbol alphabet without motivation from an underlying cluster algebra this is not something which happens in general. 

Moreover, we find that integrability plus the adjacency constraints prevents quadratic pairs of Type 3 from appearing even when separated by a single slot, as the \(D_4\) cluster algebra suggests. They may appear separated by \emph{two} in integrable quadruples, and the weight two function which sits between them is in fact unique, being the weight two function which emerges from \(\operatorname{F}(2,3;5) \cong D_4\). 

One might interpret this as evidence that the cluster structures we have described are natural to impose on the pentagon function space.   

\section{Six particle massless scattering}

We now investigate whether a similar procedure can recover information about the recently calculated two-loop, massless, six-point planar alphabet, recently presented in \cite{henn2025completefunctionspaceplanar}. Note that for \(n=6\) the flag varieties \(\operatorname{F}(2,4;n)\) and \(\operatorname{F}(2,n-2;n)\) are \emph{both} \(\operatorname{F}(2,4;6)\), whose associated cluster algebra is infinite (though of finite mutation type). However, a given symbol letter will in general have a different expression in terms of flag variables depending on whether we use the momentum twistor or spinor helicity map between flag variables and the kinematics.  

It was described in \cite{Drummond_2020} that insights from tropical geometry can provide a natural truncation of the infinite list of \(\mathcal{A}\)-coordinates generated by Gr(\(4,8\)). In particular, a truncated list of \(272\) active cluster coordinates was found to suffice to describe eight particle scattering at MHV and NMHV when supplemented with the eight frozen variables \cite{drummond2019algebraicsingularitiesscatteringamplitudes}. Of these \(272\) variables, \(71\) treat the points \(Z_7\), \(Z_8\) as a line and so have a realisation in the flag \(\operatorname{F}(2,4;6)\); we supplement these with the six frozen coordinates of Gr(\(4,8\)) which are compatible with the flag to give a list of \(77\) cluster variables. We find that this truncated list saturates the set of six point planar hexagon symbol letters which can be identified inside the flag, so that generating more \(\mathcal{A}\)-coordinates by performing many mutations does not yield anything further. 

\subsection{Six point rational letters}

Let us follow \cite{henn2025completefunctionspaceplanar} in denoting the \(289\) letters of the planar hexagon alphabet as \(\{W_i\}_{i=1}^{289}\). The algebraic letters are
\begin{equation}
\{W_i\}_{118}^{122} \cup \{W_i\}_{157}^{181}  \cup \{W_i\}_{275}^{289}
\end{equation}
and we defer an investigation of these to the next section. Note that, in order to complete the alphabet under dihedral transformations, a number of redundant letters were introduced in \cite{henn2025completefunctionspaceplanar} which are expressible as multiplicative combinations of other letters in the alphabet. These are
\begin{multline}
\{W_{i}\}_{i=191}^{193} \cup \{W_{i}\}_{i=212}^{217} \cup \{W_{i}\}_{i=221}^{223} \cup \{W_{i}\}_{i=248}^{274} \cup \{W_{i}\}_{i=279}^{280} \\ \cup \{W_{i}\}_{i=287}^{289}.
\end{multline}

The \(244\) rational letters may be separated into thirty-six cyclic classes of six and eight cyclic classes of three, whose expressions we give in the appendix. Note that here we class a letter as rational if it is expressible as a rational function in terms of spinor brackets; some of these letters do \emph{not} admit a rational expression in terms of Mandelstam invariants. We do not class letters \(\{W_{118+i}\}_{i=0}^{4}\) as rational since they correspond to square root letters \(\{r_i\}_{i=1}^5\) respectively; we would expect, and will later see, that these are associated to infinite limit rays of a cluster algebra rather than \(\mathcal{A}\)-coordinates. 

Indeed, many of these cyclic classes are themselves non-cyclic permutations of each other, and so the entire alphabet decomposes into just nineteen permutation classes as given in Table 1.

\renewcommand{\arraystretch}{1.3}
\begin{table}
  \centering
  \begin{tabular}{ll}
    \toprule
    \begin{minipage}{2cm}
      Permutation\\ class $S_i$
    \end{minipage}
    &
      \begin{minipage}{2cm}
        Letters\\ in class $S_i$
      \end{minipage}
    \\
    \midrule
    $S_1$\qquad\qquad&$\{W_i\}_{i=1}^6 \cup \{W_i\}_{i=28}^{33} \cup \{W_i\}_{i=46}^{48}$\\
    $S_2$&$\{W_i\}_{i=7}^9 $\\
    $S_3$&$\{W_i\}_{i=10}^{27} \cup \{W_i\}_{i=34}^{45}$\\
    $S_4$&$\{W_i\}_{i=49}^{51} \cup \{W_i\}_{i=76}^{99} $\\
    $S_5$&$\{W_i\}_{i=52}^{57} \cup \{W_{i}\}_{i=70}^{75}$\\
    $S_6$&$\{W_i\}_{i=58}^{69} $\\
    $S_7$&$\{W_i\}_{i=100}^{105} $\\
    $S_8$&$\{W_i\}_{i=106}^{117} $\\
    $S_9$&$\{W_i\}_{i=123}^{137} $\\
    $S_{10}$&$\{W_{138}\}$\\
    $S_{11}$&$\{W_i\}_{i=139}^{144} \cup \{W_i\}_{i=151}^{156}$\\
    $S_{12}$&$\{W_i\}_{i=145}^{150} $\\
    $S_{13}$&$\{W_i\}_{i=182}^{229} \cup \{W_i\}_{i=254}^{259}$\\
    $S_{14}$&$\{W_i\}_{i=230}^{241} $\\
    $S_{15}$&$\{W_i\}_{i=242}^{247} $\\
    $S_{16}$&$\{W_i\}_{i=248}^{253} $\\
    $S_{17}$&$\{W_i\}_{i=260}^{262} $\\
    $S_{18}$&$\{W_i\}_{i=263}^{268} $\\
    $S_{19}$&$\{W_i\}_{i=269}^{274} $\\
    \bottomrule\\
  \end{tabular}
  \caption{Permutation classes of the rational letters planar six-point alphabet. The rational letters can be grouped into 19 sets where the elements of each set are related to each other by a (potentially non-cyclic) permutation on the particle labels.}
\end{table}

Using the momentum twistor embedding of the kinematics, we find that we are able to identify a substantial portion of the rational alphabet as multiplicative combinations of the \(71\) letters from the tropically truncated list of \(\operatorname{Gr}(4,8)\) letters, plus the \(\mathcal{A}\)-coordinates associated to the six frozen nodes in Gr(\(4,8\)) which treat \(78\) as a line. Namely, we can identify \(54\) symbol letters: 
$W_{1}$--$W_9$, $W_{17}$, $W_{18}$, $W_{21}$, $W_{24}$, $W_{25}$, $W_{27}$, $W_{29}$, $W_{30}$, $W_{32}$, $W_{33}$, $W_{39}$, $W_{45}$, $W_{48}$, $W_{49}$, $W_{50}$, $W_{51}$, $W_{57}$, $W_{71}$, $W_{79}$, $W_{86}$, $W_{89}$, $W_{91}$, $W_{93}$, $W_{107}$, $W_{114}$, $W_{182}$, $W_{183}$, $W_{185}$, $W_{186}$, $W_{189}$, $W_{190}$, $W_{192}$, $W_{193}$, $W_{199}$, $W_{203}$, $W_{207}$, $W_{211}$, $W_{213}$, $W_{215}$, $W_{231}$, $W_{237}$, $W_{248}$, $W_{249}$, $W_{264}$, $W_{273}$.

If we complete these under permutations on the six particle labels, we are able to recover eleven out of the nineteen permutation classes. The classes which we fail to obtain in this way are \(S_6\), \(S_7\), \(S_9\), \(S_{10}\), \(S_{11}\), \(S_{12}\), \(S_{15}\), and \(S_{17}\). Note that the only letters which require non-cyclic permutations to obtain are \(\{W_i\}_{i=10}^{15} \cup \{W_i\}_{i=94}^{99} \cup \{W_i\}_{i=218}^{229}\). Note further that classes \(S_{10}\) and \(S_{17}\) only appear at \(\mathcal{O}(\epsilon)\) in dimensional regularisation and so would be expected to drop out of appropriately defined finite quantities; \(W_{138}\) in particular is directly analogous to \(V_{31}\) for the five point alphabet. 

Unlike at five points, we find that, at six points, using the spinor helicity embedding only allows identification of a subset of the letters which are identifiable in the momentum twistor embedding. Without using permutations we can recover a larger subset of \(75\) symbol letters, namely
$W_1$--$W_7$,   $W_{19}$, $W_{20}$, $W_{22}$, $W_{23}$, $W_{46}$ -- $W_{50}$,  $W_{76}$, $W_{77}$,  $W_{80}$, $W_{82}$, $W_{83}$, $W_{85}$,   $W_{88}$, $W_{89}$,  $W_{90}$, $W_{92}$, $W_{94}$, $W_{96}$, $W_{106}$, $W_{113}$, $W_{182}$, $W_{184}$, $W_{185}$, $W_{187}$, $W_{188}$, $W_{190}$, $W_{191}$, $W_{193}$, $W_{194}$, $W_{196}$, $W_{197}$,  $W_{200}$, $W_{202}$, $W_{205}$, $W_{206}$, $W_{208}$, $W_{210}$, $W_{211}$, $W_{212}$, 
 $W_{214}$, $W_{215}$, $W_{216}$, $W_{218}$, $W_{220}$, $W_{221}$, $W_{224}$, $W_{226}$, 
 $W_{230}$, $W_{238}$, $W_{248}$, $W_{249}$, $W_{250}$, $W_{251}$, $W_{252}$, $W_{253}$, 
 $W_{255}$, $W_{257}$, $W_{263}$, $W_{274}$

but we do not find a representative of any additional permutation class, and in fact can no longer recover any representative of \(S_5\). Interestingly, this is a class which is absent from the hexagonal Wilson loop with a Lagrangian operator insertion \cite{carrôlo2025hexagonalwilsonlooplagrangian}, which is the only two-loop six-point data presently available. The only obtained letters requiring non-cyclic permutations are now \(\{W_i\}_{i=10}^{15} \cup \{W_i\}_{i=34}^{45}\).

Going beyond the tropically truncated list of \(272\) Gr\((4,8)\) letters, e.g. by collecting a large number of \(\mathcal{A}\)-coordinates by performing many mutations from the initial cluster, does not enable the recovery of any of the missing classes of symbol letters. In principle one might also hope that an unobtainable letter might non-cyclically permute to something which is outside of the planar alphabet but which can be identified in the flag, or that an unobtainable letter might originate as a product of cluster variables coming from different orientations of the flag; however, this seems not to be the case for these missing letters which we have been unable to identify as (products of) cluster variables. 

In the ancillary files, for each symbol letter obtainable under either embedding of the kinematics into flag variables we give a list of all permutations under which that letter becomes identifiable in the flag, and for each permutation give the resulting expression in terms of the truncated list of Gr(\(4,8\)) variables. A separate file is given for each of the momentum twistor and spinor helicity embeddings of the kinematics.

\subsubsection*{Comment on unused flag variables}

Note that not all of the variables on the truncated list of \(71 + 6\) \(\operatorname{Gr}(4,8)\) variables are actually needed in the momentum twistor embedding; twenty-two are not used identifying any symbol letter under any permutation, although in many cases they clearly would be used if going beyond the truncated list. For instance, the variable \(p_{1378}\) is not used while \(p_{2478}\) is; despite the fact these two variables are related to each other by a (cyclic) permutation on particle labels, \(p_{1378}\) is not used on account of the fact that the cyclic copies of the \emph{other} variables \(p_{2378}\) combines with to form symbol letters are missing from the truncated list. 

For the spinor helicity embedding, all of the cluster variables are used in some permutation.

\subsection{Six point algebraic letters}

\subsubsection{Algebraic letters in the planar hexagon alphabet}
The six point planar alphabet contains \(45\) algebraic letters containing five distinct square roots \(r_1\) to \(r_5\). Five of the algebraic letters are these naked discriminants.

The square roots are given by 
\begin{multline}
    r_{1+i} \equiv \\
C^i \bigl(\sqrt{s_{12}^2 + s_{34}^2 + s_{56}^2 - 2s_{12}s_{34}-2s_{12}s_{56}-2s_{34}s_{56}}\bigr) \\ i = 0,1
\end{multline}
and
\begin{equation}
r_{3+i} \equiv C^i \bigl(\sqrt{(s_{123} + s_{345})^2 - 4s_{12}s_{45}}\bigr) \hspace{1cm} i =0,1,2.
\end{equation}

where \(C^i\) denotes cycling forwards by one on the particle labels. Note that all five \(r_i\) are related by (in general, non-cyclic) permutations.

\subsubsection{A review of \(\operatorname{Gr}(4,8)\) origin clusters and limit rays}
Although Gr(\(4,8\)) is an infinite cluster algebra, it is of finite mutation type, with infinitely many \(\mathcal{A}\)-coordinates specifically arising from \emph{origin clusters}. An example of an origin cluster is shown in Figure \ref{fig:origin}.

\begin{figure}
  \centering
  \begin{tikzpicture}
    % \node (b1) at (1,0) {$b_1$};
    \node (b2) at (-1,0) {$b_2$};
    \node (b3) at (-2,0) {$b_3$};

    \node (z0) at (0,1) {$z_0$};
    \node (w0) at (0,-1) {$w_0$};

    \node (b1) at (1,0) {$b_1$};

    \node (a4) at (2,0) {$a_4$}; 
    \node (a3) at (3,0) {$a_3$};
    \node (a2) at (-3,0) {$a_2$}; 
    \node (a1) at (-4,0) {$a_1$};

    \draw[-latex] ($(w0) + (0.1,0.2)$) -- ($(z0) + (0.1,-0.2)$);
    \draw[-latex] ($(w0) + (-0.1,0.2)$) -- ($(z0) + (-0.1,-0.2)$);
    
    \draw[-latex] (z0) -- (b3);
    \draw[-latex] (b3) -- (w0);
    \draw[-latex] (z0) -- (b2);
    \draw[-latex] (b2) -- (w0);
    \draw[-latex] (z0) -- (b1);
    \draw[-latex] (b1) -- (w0);

    \draw[-latex] (a2) -- (b3);
    \draw[-latex] (a1) -- (a2);
    \draw[-latex] (a4) -- (a3);
    \draw[-latex] (b1) -- (a4);

    \node (nodevalues1) at (-2.3,-3) {
      $\displaystyle
      \begin{aligned}[t]
        z_0 &=p_{1236}\, p_{1578} - p_{1235}\, p_{1678}\\
        b_1 &= p_{1236}\, p_{3578} - p_{1235}\, p_{3678}\\
        b_2 &= p_{1256}\\
              b_3 &= p_{1346}\,p_{1578} - p_{1345}\,p_{1678}\\
        a_4 &=p_{1236}\, p_{4578} - p_{1235}\, p_{4678}+ p_{1234}\, p_{5678}\\
      \end{aligned}
      $      
    };
    \node (nodevalues1) at (2,-3) {
      $\displaystyle
      \begin{aligned}[t]
        w_0 & =p_{1356}\\
        a_1 &= p_{1345}\\
        a_2 &= p_{1346}\\
        a_3 &= p_{1237}\\
       \phantom{ a_4} &\phantom{= p_{1237}}\\
      \end{aligned}
      $      
    };
    
  \end{tikzpicture}
  \caption{One of the 64 origin clusters associated with discriminant $\Delta$ that appears in 8-point amplitudes in MSYM and six point Wilson loop correlator with a local operator. The nodes labelled by $a_1$, $a_2$, $a_3$, and $a_4$ do not enter the algebraic letters produced by the infinite mutation sequence starting by this origin cluster, and they can be mutated arbitrarily. We choose to present here the representative that is also a flag cluster. The frozen nodes are not displayed.}
  \label{fig:origin}
\end{figure}

In \cite{drummond2019algebraicsingularitiesscatteringamplitudes}, it was noted that each origin cluster supplies an infinite sequence of \(\mathcal{A}\)-coordinates by repeated mutation on the nodes with variables \(z_0\) and \(w_0\). Collecting the \(b\) variables together as \(b = b_1b_2b_3\), and similarly combining the outgoing frozen variables from \(w_0\) as \(f_w\) and the incoming frozen variables to \(z_0\) as \(f_z\), it is shown in {\it loc.cit.} that the cluster \(\mathcal{A}\)-coordinates which are generated obey the recursion relations
\begin{equation}
z_{n+2}z_n = b\,f_z\,\mathcal{F}^n + z_{n+1}^2
\end{equation}
and
\begin{equation}
w_{n+2}w_n = b\,f_w\,\mathcal{F}^n + w_{n+1}^2\,.
\end{equation}
As such we have two separate infinite sequences depending on whether we first mutate on \(w_0\) or \(z_0\). Here we define 
\begin{equation}
\mathcal{F} = f_wf_z
\end{equation}
as the product of the frozen coordinates.

It is shown in \cite{drummond2019algebraicsingularitiesscatteringamplitudes} that this recursion relation for the \(z\)'s is solved by 
\begin{multline}
z_n = \frac{1}{2^{n+1}}\bigl((z_0 + \mathcal{B}_z\sqrt{\Delta})(\mathcal{P}_z+\sqrt{\Delta})^n \\+ (z_0 - \mathcal{B}_z\sqrt{\Delta})(\mathcal{P}_z - \sqrt{\Delta})^n\bigr)
\end{multline}
where
\begin{align}
  \mathcal{P}_z = \frac{f_zw_0 + z_1}{z_0},\qquad \mathcal{B}_z = \frac{2z_1 - z_0 \mathcal{P}_z}{\Delta},\nonumber\\
\text{and}\qquad  \Delta = \mathcal{P}_z^2 - 4\mathcal{F}_z\,.
\end{align}

The solution of the recurrence relation for the \(w_n\) and $z_n$ is solved by taking the above formulae and swapping \(z\) with \(w\); note that it turns out \(\mathcal{P}_z = \mathcal{P}_w\).

If one repeatedly applies the mutation rules on the \(\textbf{g}\)-vectors associated to the \(\mathcal{A}\)-coordinates \(z_n\) and \(w_n\), the \(\textbf{g}\)-vectors asymptote to the same \emph{limit ray} \(\textbf{g}_{\infty}\), and any origin cluster with the same \(\Delta\) will have the same limit ray. It is therefore natural to associate a single square root \(\sqrt{\Delta}\) to each limit ray. Moreover, since it is observed in \(\operatorname{Gr}(4,8)\) that the product
\begin{equation}
(z+\mathcal{B}_z\sqrt{\Delta})(z-\mathcal{B}_z\sqrt{\Delta}) 
\end{equation}
is always a product of cluster variables divided by \(\Delta\), it is natural to consider 
\begin{equation}
\phi_z = \frac{z_0 + \mathcal{B}_z\sqrt{\Delta}}{z_0 - \mathcal{B}_z\sqrt{\Delta}}
\end{equation}
and
\begin{equation}
\phi_w = \frac{w_0 + \mathcal{B}_w\sqrt{\Delta}}{w_0 - \mathcal{B}_w\sqrt{\Delta}}
\end{equation}
as well as \(\Delta\) itself as new, algebraic letters associated to each origin cluster. 

In the case of \(\operatorname{Gr}(4,8)\), there are infinitely many origin clusters with infinitely many different limit rays; in \cite{drummond2019algebraicsingularitiesscatteringamplitudes}, a truncation is used which gives two square roots and \(64\) origin clusters. These \(64\) origin clusters supply \(128\) algebraic letters, of which \(18\) are multiplicatively independent and span precisely the \(18\) algebraic letters in the two-loop NMHV octagon amplitude in MSYM. 

Given the realisation of \(\operatorname{F}(2,4;6)\) as a codimension two subalgebra of \(\operatorname{Gr}(4,8)\), it is natural to ask to what extent the same procedure in the flag can generate the algebraic letters in the six point alphabet. 

\subsubsection{Algebraic letters from origin clusters in \(\operatorname{F}(2,4;6)\)}
The case of \(\operatorname{F}(2,4;6)\) is very similar to that of \(\operatorname{Gr}(4,8)\) given that the former can be embedded in the latter as a cluster subalgebra. The only difference is that there are now only \(16\) different origin clusters, each of which asymptotes to the same ray. We are thus supplied a single square root letter

\begin{multline}
\Delta =p_{12}^2 p_{3456}^2 - 
 2 p_{1256}\, p_{12} p_{3456}\, p_{34} + 
 p_{1256}^2 p_{34}^2 \\ - 
 2 p_{1234}\, p_{12} p_{3456}\, p_{56} - 
 2 p_{1234}\, p_{1256} p_{3478}\, p_{56} + 
 p_{1234}^2 p_{56}^2\,.
\end{multline}

This is precisely \(r_1^2\) when evaluated in the spinor helicity embedding, or (up to some simple factors of squared cluster variables) \(r_2^2\) when evaluated in the momentum twistor embedding. It is striking that the two embeddings of the kinematics conspire to supply the same square roots after permutation completion; even though in general a function of Plücker coordinates would have a very different kinematic meaning depending on the interpretation in terms of momentum twistors or spinor helicity variables, the single square root supplied by the flag essentially agrees up to cycling in either interpretation. Clearly upon completing under permutations, we thus recover all of the square roots \(r_1\) through \(r_5\) in either embedding of the kinematics (as well as many other square roots which are not relevant for the planar alphabet).

Each of the sixteen origin clusters supplies two algebraic letters via the construction of \(\phi_z\) and \(\phi_w\), and it is straightforward to compute that seven of these thirty-two algebraic letters are multiplicatively independent. Within the span of these seven, we can obtain all of the algebraic letters involving \(r_1\) using the spinor helicity embedding of the kinematics, and all of the algebraic letters involving \(r_2\) using the momentum twistor embedding of the kinematics (with the exception of the naked square roots themselves, which are already identified via the product \((z+\mathcal{B}_z\sqrt{\Delta})(z-\mathcal{B}_z\sqrt{\Delta})\) associated to each origin cluster).

Completing either set under permutations on particle labels generates every algebraic letter in the alphabet. 

\subsection{Comments on adjacency relations for the hexagonal Wilson loop}
While we defer any detailed investigation of adjacency relations at six points to future work, let us briefly remark that the quadratic letters \(\{V_i\}_{i=6}^{10}\) from the pentagon alphabet are equivalent to \(\{W_{i}\}_{i=10}^{15}\) in the hexagon alphabet, while \(\{V_i\}_{i=11}^{15}\) from the pentagon alphabet are equivalent to the hexagon letters \(\{W_{i}\}_{i=22}^{27}\). It is interesting to ask whether the observations regarding adjacencies and triples from the five-point case now carry over to the recently calculated planar hexagonal Wilson loop with operator insertion. 

Let us first note that not all of the rational symbol letters appearing in the planar hexagonal Wilson loop with Lagrangian operator insertion, recently presented in \cite{carrôlo2025hexagonalwilsonlooplagrangian}, can be generated using either the momentum twistor or spinor helicity embeddings. In particular, we fail to reproduce the permutation classes \(S_6\), \(S_7\) and \(S_{14}\) which do feature e.g. through the cyclic class containing \(W_{58}\). 

Empirically, the only six-point letters we have identified here as analogous to the five-point quadratics which now appear adjacently in the symbol are \(W_{22} \otimes W_{25}\) and its cyclic permutations. There are then \emph{two} cyclic classes of quadratics which may appear only when separated by one slot, and in each case the slot between is inhabited by a unique combination:

\begin{enumerate}
\item \(W_{23} \otimes \frac{W_1 W_7}{W_2W_{28}} \otimes W_{10}\) and cyclic permutations. This class seems to have a realisation in the flag, specifically using the momentum twistor embedding. For example, if one interchanges the particle labels under the permutation \(\{2,4,3,1,5,6\}\), \(W_{23}\) has a factor which mutates to one of \(W_{10}\)'s factors, and the \(\mathcal{X}\)-coordinate of \(W_{23}\)'s factor is precisely \( \frac{W_1 W_7}{W_2W_{28}}\) subjected to that same permutation and interpreted using the momentum twistor embedding of the kinematics. We could \emph{not} find a cluster which realises this triple for any permutation in the spinor helicity embedding; in particular, there is no permutation on particle labels such that all six letters involved are mutually identifiable in terms of our truncated list of flag variables.

\item \(W_{26} \otimes \frac{W_1 W_{31}}{W_{61}} \otimes W_{22}\) and cyclic permutations. Since \(W_{61}\) is one of the letters which we fail to reproduce in either embedding of the kinematics, we cannot hope to realise this inside \(\operatorname{F}(2,4;6)\). However, it is interesting to note that these letters do obey the exchange-type relation
\begin{equation}
W_{26}W_{22} = W_1W_{31} - W_{61}
\end{equation}
which is indicative that there may be some cluster structure involved even for those symbol letters which we presently fail to obtain. 
\end{enumerate}

\section{Further comments on observables}

We now briefly comment on some further observations which can be made about the available data.  

\subsection{Pentagonal Wilson loop - Lagrangian correlator}
As already noted, the terms in the pentagonal Wilson loop with Lagrangian insertion \cite{chicherin2022pentagonwilsonlooplagrangian} which contain two quadratic letters separated by one are highly indicative of an \(\operatorname{F}(2,4;5) \cong A_4\) interpretation. It is therefore natural to wonder if the coefficients of the leading singularities, \(r_0\) through \(r_5\), can be expressed purely in terms of \(A_4\) cluster-adjacent polylogarithms. Note that this provides a stronger constraint on integrable words than e.g. purely imposing that 'quadratic' pentagon letters do not appear adjacently. In particular, there are \(370\) integrable, homogeneous weight four words from the planar alphabet with good initial entries which do not include distinct quadratics appearing adjacently, while there are only \(339\) such words which are obtained as permutations of the weight four cluster adjacent polylogarithms arising from the canonical orientation of the flag. 

In particular, we performed the following exercise:
\begin{enumerate}
\item Take homogeneous combinations of the cluster variables associated to the partial flag variety \(\operatorname{F}(2,4;5)\), and construct all weight-four integrable words which obey \(A_4\) cluster adjacency. Express these in terms of the letters of the planar pentagon alphabet. 
\item Complete the weight-four integrable words under arbitrary permutations on the particle momenta, to obtain a complete basis of cluster-adjacent weight four integrable words coming from any orientation of \(\operatorname{F}(2,4;5)\). 
\item Test whether the symbols of the coefficients of the Wilson loop correlator may be expressed as a linear combination of these basis elements. 
\end{enumerate}

Doing so, we \emph{were} able to express the symbols of the coefficients of the leading singularities for the pentagonal Wilson loop with Lagrangian operator insertion using \(A_4\)-type contributions only. 

\subsection{Form factors}
The MHV four-particle form factor for the chiral part of the stress-tensor supermultiplet in planar MSYM was recently calculated at three loops in \cite{Dixon:2022xqh}. Due to the normalisation by a massive operator \(q^2 \cong (p_5 + p_6)\), the symbol letters naturally admit a representation using six particle labels, and so it is interesting to test the relevance of \(\operatorname{F}(2,4;6)\) in describing this observable. 

The symbol alphabet consists of \(52\) rational letters, plus many algebraic letters which we do not attempt to analyse in detail here. In the language of \cite{Dixon:2022xqh}, where the alphabet is given, the rational letters are
\begin{equation}
\{\mathfrak{a}_i\}_{i=1}^{44} \cup \{\mathfrak{a}_i\}_{i=47}^{54}
\end{equation}
Note that although five other symbol letters are rational in terms of Mandelstam invariants, we do not consider them rational letters since they are in fact the square roots in the alphabet and thus would be expected to come from limit rays rather than  cluster variables. 

Using the momentum twistor embedding of the kinematics, after completing under permutations on the particle labels we recover every rational letter with the exception of the permutation class \(\{\mathfrak{a}_{i}\}_{i=35}^{42}\) as a multiplicative combination of \(\operatorname{Gr}(4,8)\) cluster variables from the tropically truncated list of \(272\) \(\mathcal{A}\)-coordinates. Using the spinor helicity embedding fails to recover more rational letters, namely \(\{\mathfrak{a}_{i}\}_{i=27}^{34}\) in addition to those missed using momentum twistors. 

Using the larger tropically truncated list of \(544\) \(\operatorname{Gr}(4,8)\) \(\mathcal{A}\)-coordinates given in \cite{Drummond_2020} does not allow us to identify any of the missing letters, but by mutating beyond even this larger truncated list, we can in fact recover the missing class of rational letters using the momentum twistor embedding of the kinematics. For instance, after subjecting  \(\mathfrak{a}_{35}\) to the permutation \((2,3,4,5,1,6)\) on particle labels, it matches (up to factors of simpler cluster variables from the truncated list) the \(\operatorname{Gr}(4,8)\) cluster variable with vector
\begin{equation}
(2, -1, 0, -2, 0, 2, 0, 2, -2).
\end{equation}
in the tropical fan, which is a generalisation of the tropical Grassmannian construction of Speyer and Williams \cite{speyer2003tropicaltotallypositivegrassmannian} based on all Pl\"ucker coordinates and their parity images which include quadratic letters. The other rational letters missed in in the momentum twistor embedding are all permutations of this letter, and so the entire rational alphabet is recovered. 

Note that it is not entirely surprising that the form factor requires \(\operatorname{Gr}(4,8)\) letters beyond those relevant for six particle massless scattering, since it receives contributions from integrals with a non-planar topology.  Going beyond the truncated list of letters does \emph{not} enable the identification of any further letters in the spinor helicity embedding of the kinematics. 

Since the three-loop symbol is of weight six, it is also an instructive exercise to infer empirical adjacency relations from the data (i.e. simply observe which symbol letters never appear next to each other) and construct all weight-three integrable words obeying these constraints. One then finds \(64\) fully rational triples where two letters we have forbidden to appear adjacently are separated by one slot with a unique inhabitant. Eight of these \(a \otimes X \otimes a'\) triples do not obey an exchange-type relation of the form 
\begin{equation}
aa' = \textrm{Numerator}(X) \pm \textrm{Denominator}(X)
\end{equation}
and so are presumably accidental, while in every other case (including those involving \(\mathfrak{a}_{27}\) and its permutation copies, which requires going beyond the truncated list of \(\mathcal{A}\)-coordinates) we are able to find an explanatory cluster for that triple. More precisely, for the triple \(a \otimes X \otimes a'\) and using the momentum twistor embedding of the kinematics we can find a permutation on the particle labels such that there exists a cluster where a factor of \(a\) mutates to a factor of \(a'\) and the \(\mathcal{X}\)-coordinate associated to that mutation is precisely \emph{the} \(X\) which sits between the two letters. 

\section{Conclusions}

Although we have presented evidence that partial flag varieties contain information relevant for non-dual conformal scattering at five and six points, this is not a complete story. For five-point massless scattering, there is already evidence \cite{liu2025analyticcomputationthreeloopfivepoint} that more symbol letters appear at three loops, which we do not obtain here using either embedding of the kinematics. Already at two loops we fail to obtain a number of six point symbol letters, including many e.g. \(\{W_{i}\}_{i=58}^{63}\) which have appeared in the hexagonal Wilson loop with Lagrangian operator insertion. Although we have noted that there is some evidence of a cluster structure underpinning even these letters, it is not clear what this ought to be, nor what the cluster structure available for the quadratic triples observed in the non-planar data might be. 

It is striking is that two completely different embeddings of the kinematics inside (different, for \(n \neq 6\)) partial flag varieties both show success at generating large portions of the relevant symbol alphabets. Given that beyond five points the momentum twistor embedding seems to generate more symbol letters successfully (and in particular that the spinor helicity embedding always seems to generate a subset of those obtainable with momentum twistors), and seems to better align with the simple adjacency relations observed in the data so far, a question which remains in need of resolution is whether \(\operatorname{F}(2,n-4;n)\) with the spinor helicity embedding is ever needed to generate terms without a sound \(\operatorname{F}(2,4;n)\), momentum twistor interpretation. 

A particularly pressing task which remains is clearly to understand what cluster algebra interpretation, if any, can be given to the missing two-loop six point letters. While some of the missing classes only feature at \(\mathcal{O}(\epsilon)\) in dimensional regularisation, and others might yet drop out of appropriately defined finite quantities in the manner seen for the pentagon letter \(V_{31}\), the letters \(\{W_i\}_{i=58}^{69}\), \(\{W_i\}_{i=100}^{105}\) and \(\{W_i\}_{i=242}^{247}\) have already appeared in the hexagonal Wilson loop with Lagrangian operator insertion at two loops.

Given the use of \(\operatorname{Gr}(4,8)\), namely the tropically truncated list of \(272\) letters, combined with a massless limit in \cite{Chicherin_2021} to reproduce the two-loop planar pentagon alphabet using cyclic permutations only, it would be interesting to investigate whether any different truncations of the \(\operatorname{Gr}(4,8)\) alphabet could yield candidates for the missing three-loop pentagon letters, and whether the square roots being reported there could be identified as infinite limit rays of \(\operatorname{F}(2,4;6)\). Likewise, following a similar procedure to embed six particle kinematics in \(\operatorname{Gr}(4,9)\) would be a natural avenue to pursue in seeking the missing six point letters.

If a more complete understanding can be achieved of the relevance of partial flag varieties (and indeed any other cluster algebras) to observables without dual conformal symmetry, constraints on the analytic structure may prove powerful for symbol bootstrap calculations. Given the extraordinary success of this approach at high loop-order in MSYM theory, it would be of particular note if this technology could ultimately be brought to bear on phenomenologically relevant processes e.g. QCD. 
\appendix
\section{Rational Letters in the Six Point Planar Hexagon Alphabet}
In this appendix we include Table \ref{tab:sixalphabet} listing the cyclic classes of letters of the two-loop six point alphabet of \cite{henn2025completefunctionspaceplanar}. We also indicate which letters appear in the correlator of the hexagonal Wilson loop with a local operator \cite{carrôlo2025hexagonalwilsonlooplagrangian}, and whether these letters can be recovered from the momentum-twistor flag embedding of the kinematics into $\operatorname{Gr}(4,8)$.
\newcommand{\hexagon}{\mathord{\raisebox{0.6pt}{\tikz{\node[draw,scale=.65,regular polygon, regular polygon sides=6,fill=none](){};}}}}
\newcommand{\hexcheck}{$\hexagon$\hspace{-2mm}\raisebox{0.75mm}{\checkmark}}
\begin{widetext}
  \begin{longtable}{lll}
    \toprule
    Letter name & Expression in terms of momenta&\checkmark\\
    \midrule
$\displaystyle W_{1} $&$\displaystyle s_{12}$&\hexcheck\\
$\displaystyle W_{7} $&$\displaystyle s_{123}$&\hexcheck\\
$\displaystyle W_{10} $&$\displaystyle s_{12}+s_{23}$&\hexcheck\\
$\displaystyle W_{16} $&$\displaystyle s_{12}-s_{123}$&\hexcheck\\
$\displaystyle W_{22} $&$\displaystyle s_{12}-s_{345}$&\hexcheck\\
$\displaystyle W_{28} $&$\displaystyle s_{123}-s_{12}-s_{23}$&\hexcheck\\
$\displaystyle W_{34} $&$\displaystyle s_{12} - s_{34} + s_{123}$&\hexcheck\\
$\displaystyle W_{40} $&$\displaystyle s_{12} - s_{56} + s_{345}$&\hexcheck\\
$\displaystyle W_{46} $&$\displaystyle s_{12} + s_{45} - s_{123} - s_{345}$&\hexcheck\\
$\displaystyle W_{49} $&$\displaystyle  -s_{12}s_{45} + s_{123}s_{345}$&\hexcheck\\
$\displaystyle W_{52} $&$\displaystyle s_{12}s_{56} - s_{12}s_{123}+s_{34}s_{123}$&$\checkmark$\\
$\displaystyle W_{58} $&$\displaystyle  -s_{12}s_{45} - s_{23}s_{234}+s_{123}s_{345}$&$\hexagon$\\
$\displaystyle W_{64} $&$\displaystyle  -s_{12}s_{45} - s_{34}s_{123}+s_{123}s_{345}$&$\hexagon$\\
$\displaystyle W_{70} $&$\displaystyle s_{12}s_{56} + s_{34}s_{123}-s_{56}s_{123}$&\checkmark\\
$\displaystyle W_{76} $&$\displaystyle s_{12}s_{34} + s_{12}s_{45}-s_{34}s_{345} + s_{56}s_{345}-s_{123}s_{345}$&\hexcheck\\
$\displaystyle W_{82} $&$\displaystyle s_{12}s_{45} + s_{23}s_{45}+s_{16}s_{123} - s_{23}s_{123} - s_{123}s_{345}$&\hexcheck\\
$\displaystyle W_{88} $&$\displaystyle  -s_{12}s_{34} + s_{12} s_{345} + s_{34} s_{345} - s_{56}s_{345} - s_{345}^2$&\hexcheck\\
$\displaystyle W_{94} $&$\displaystyle  -s_{12}^2+2s_{12}s_{34}-s_{34}^2-s_{23}s_{56}+s_{12}s_{123}-s_{34}s_{123}-s_{12}s_{234}+s_{34}s_{234}+s_{123}s_{234}$&\checkmark\\
$\displaystyle W_{100} $&$\displaystyle  -s_{12}s_{16}s_{45}-s_{23}s_{34}s_{56}-s_{16}s_{34}s_{123} +s_{12}s_{45}s_{234}+s_{34}s_{123}s_{234}+s_{23}s_{56}s_{345}+s_{16}s_{123}s_{345}-s_{123}s_{234}s_{345}$&$\hexagon$\\
$\displaystyle W_{106} $&$\displaystyle  -s_{12}s_{23}s_{56}-s_{12}s_{56}^2-s_{34}s_{56}s_{123}+s_{56}^2s_{123}+s_{12}s_{56}s_{234}+s_{34}s_{123}s_{234}-s_{56}s_{123}s_{234}$&\checkmark\\
$\displaystyle W_{112} $&$\displaystyle  -s_{16}s_{23}^2-s_{16}s_{23}s_{56}+s_{16}s_{23}s_{123}+s_{23}^2s_{234}-s_{23}s_{45}s_{234}-s_{23}s_{123}s_{234}+s_{45}s_{123}s_{234}$&\checkmark\\
$\displaystyle W_{123} $&$\displaystyle  \epsilon_{1234} $&\\
$\displaystyle W_{129} $&$\displaystyle  \epsilon_{1235} $&\\
$\displaystyle W_{135} $&$\displaystyle  \epsilon_{1245} $&\\
$\displaystyle W_{138} $&$\displaystyle  \langle12\rangle[23]\langle34\rangle[45]\langle56\rangle[61] - [12]\langle23\rangle[34]\langle45\rangle[56]\langle61\rangle$&\\
$\displaystyle W_{139} $&$\displaystyle  -\epsilon_{4561}s_{12}+\epsilon_{5612}s_{123} $&\\
$\displaystyle W_{145} $&$\displaystyle  -\epsilon_{1245}s_{34}+\epsilon_{1234}s_{56} + \epsilon_{3451}(s_{34}-s_{234}) - \epsilon_{2345}(s_{56}-s_{234})  $&\\
    $\displaystyle W_{151} $&$\displaystyle  \epsilon_{1234}s_{16}+\epsilon_{6123}s_{234} $&\\
$\displaystyle W_{182} $&$\displaystyle  \frac{-\epsilon_{1234}+s_{12}s_{23}-s_{23}s_{34}+s_{23}s_{56}+s_{34}s_{123}-s_{12}s_{234}-s_{123}s_{234})}{\epsilon_{1234}+s_{12}s_{23}-s_{23}s_{34}+s_{23}s_{56}+s_{34}s_{123}-s_{12}s_{234}-s_{123}s_{234}} $&\hexcheck\\[10pt]
$\displaystyle W_{188} $&$\displaystyle  \frac{-\epsilon_{1234}+s_{12}s_{23}-s_{23}s_{34}-s_{23}s_{56}+s_{34}s_{123}-s_{12}s_{234}+s_{123}s_{234}}{\epsilon_{1234}+s_{12}s_{23}-s_{23}s_{34}-s_{23}s_{56}+s_{34}s_{123}-s_{12}s_{234}+s_{123}s_{234}} $&\hexcheck\\[10pt]
$\displaystyle W_{194} $&$\displaystyle  \frac{-\epsilon_{1234}-s_{12}s_{23}+s_{23}s_{34}+s_{23}s_{56}+s_{34}s_{123}+s_{12}s_{234}-s_{123}s_{234}}{\epsilon_{1234}-s_{12}s_{23}+s_{23}s_{34}+s_{23}s_{56}+s_{34}s_{123}+s_{12}s_{234}-s_{123}s_{234}} $&\hexcheck\\[10pt]
$\displaystyle W_{200}  $&$\displaystyle  \frac{-\epsilon_{1234}+s_{12}s_{23}-s_{23}s_{34}+ s_{23}s_{56}+s_{34}s_{123}+s_{12}s_{234}- s_{123}s_{234}}{ \epsilon_{1234}+s_{12}s_{23}-s_{23}s_{34}+ s_{23}s_{56}+s_{34}s_{123}+s_{12}s_{234}- s_{123}s_{234}} $&\hexcheck\\[10pt]
$\displaystyle W_{206} $&$\displaystyle  \frac{-\epsilon_{1234}+s_{12}s_{23}-s_{23}s_{34}-s_{23}s_{56}+s_{34}s_{123}+s_{12}s_{234}+2s_{23}s_{234}-s_{123}s_{234}}{\epsilon_{1234}+s_{12}s_{23}-s_{23}s_{34}-s_{23}s_{56}+s_{34}s_{123}+s_{12}s_{234}+2s_{23}s_{234}-s_{123}s_{234}} $&\hexcheck\\[10pt]
$\displaystyle W_{218} $&$\displaystyle  \frac{-\epsilon_{1234}+s_{12}s_{23}-s_{23}s_{34}+2s_{12}s_{56}+s_{23}s_{56}-2s_{12}s_{123}+s_{34}s_{123}-s_{12}s_{234}-s_{123}s_{234}}{\epsilon_{1234}+s_{12}s_{23}-s_{23}s_{34}+2s_{12}s_{56}+s_{23}s_{56}-2s_{12}s_{123}+s_{34}s_{123}-s_{12}s_{234}-s_{123}s_{234}} $&\hexcheck\\[10pt]
$\displaystyle W_{224} $&$\displaystyle  \frac{-\epsilon_{1234}+2s_{12}^2+s_{12}s_{23}-2s_{12}s_{34}-s_{23}s_{34}+s_{23}s_{56}-2s_{12}s_{123}+s_{34}s_{123}+s_{12}s_{234}-s_{123}s_{234}}{\epsilon_{1234}+2s_{12}^2+s_{12}s_{23}-2s_{12}s_{34}-s_{23}s_{34}+s_{23}s_{56}-2s_{12}s_{123}+s_{34}s_{123}+s_{12}s_{234}-s_{123}s_{234}} $&\checkmark\\[10pt]
$\displaystyle W_{230} $&$\displaystyle  \frac{\epsilon_{1234}+s_{12}s_{23}-s_{23}s_{34}+2s_{12}s_{56}+s_{23}s_{56}+s_{34}s_{123}-2s_{56}s_{123}-s_{12}s_{234}+s_{123}s_{234}}{-\epsilon_{1234}+s_{12}s_{23}-s_{23}s_{34}+2s_{12}s_{56}+s_{23}s_{56}+s_{34}s_{123}-2s_{56}s_{123}-s_{12}s_{234}+s_{123}s_{234}} $&\checkmark\\[10pt]
$\displaystyle W_{236} $&$\displaystyle  \frac{\epsilon_{1234}-s_{12}s_{23}+s_{23}s_{34}+s_{23}s_{56}+2s_{34}s_{56}-s_{34}s_{123}+s_{12}s_{234}-2s_{56}s_{234}+s_{123}s_{234}}{-\epsilon_{1234}-s_{12}s_{23}+s_{23}s_{34}+s_{23}s_{56}+2s_{34}s_{56}-s_{34}s_{123}+s_{12}s_{234}-2s_{56}s_{234}+s_{123}s_{234}} $&\checkmark\\[10pt]
$\displaystyle W_{242} $&$\displaystyle  \frac{-\epsilon_{1235}-s_{12}(s_{16}+s_{45}-s_{234})+s_{23}(s_{34}+s_{56}-s_{345})+s_{123}(s_{16}-s_{34}-s_{234}+s_{345})}{\epsilon_{1235}-s_{12}(s_{16}+s_{45}-s_{234})+s_{23}(s_{34}+s_{56}-s_{345})+s_{123}(s_{16}-s_{34}-s_{234}+s_{345})} $&$\hexagon$\\[10pt]
\bottomrule\\
      \caption{All rational letters in the planar two-loop six-point alphabet presented in \cite{}. The remainder of the letters can be obtained as cyclic copies of those listed here. We omitted those letters not multiplicatively independent of these. The table is annotated according to whether the momentum-twistor embedding captures the letter by a checkmark (\checkmark). If the letter appears in the two-loop correlator of a hexagonal Wilson loop with a local operator in MSYM we indicated this with a hexagon symbol.} \\
  \label{tab:sixalphabet}\\

\end{longtable}
\end{widetext}

\section*{Acknowledgements}
The authors thank Jianrong Li for helpful discussions, and Marcus Spradlin for discussions at the ``Summer School in Total Positivity and Quantum Field Theory'' at CMSA, Harvard where we became aware of their overlapping work. JMD, \"OCG, and RW are supported by the STFC consolidated grant ST/X000583/1. \"OCG is also supported by the Royal Society University Research Fellowship URF\textbackslash R1\textbackslash221236.
LB is supported by PAPIIT IA100724 UNAM dgapa 2024 and SECIHTI CF-2023-G-106.
\bibliography{six-letter}{} \bibliographystyle{unsrtnat}
\end{document}